\documentclass[12pt]{article}
\usepackage[cp1251]{inputenc}

\usepackage{indentfirst}
\usepackage{graphicx}
\usepackage{subfigure}
\usepackage{amssymb}

\begin{document}

\pdfoutput=1


\title{Anomalies of Density, Stresses, and the Gravitational Field
in the Interior of Mars}

\author{N.A. Chuikova\thanks{chujkova@sai.msu.ru}, L.P. Nasonova
\thanks{nason@sai.msu.ru}, and T. G. Maximova \\
Lomonosov Moscow State University,  \\ Sternberg State Astronomical Institute, \\
{\small  Universitetskii prospect, 13, Moscow, 119991, Russia}}
\date{}

\maketitle
\medskip


\begin{abstract}
We determined the possible compensation depths for relief harmonics of different degrees and
orders. The relief is shown to be completely compensated within the depth range of 0 to 1400 km. The lateral
distributions of compensation masses are determined at these depths and the maps are constructed. The possible nonisostatic vertical stresses in the crust and mantle of Mars are estimated to be 64 MPa in compression
and 20 MPa in tension. The relief anomalies of the Tharsis volcanic plateau and symmetric feature in the eastern hemisphere could have arisen and been maintained dynamically due to two plumes in the mantle substance that are enriched with fluids. The plumes that originate at the core of Mars can arise and be maintained
by the anomalies of the inner gravitational field achieving +800 mGal in the region of plume formation, –
1200 mGal above the lower mantle–core transition layer, and –1400 mGal at the crust

\end{abstract}

\bigskip
DOI: 10.3103/S0027134912020075


\newpage

{\Large \bf \quad Introduction }

\bigskip

\qquad Since seismic studies have not been carried out on
Mars, preliminary information about its internal composition can be obtained using the observational data on its gravitational field and relief, as well as some theoretical conclusions verified when studying the density structure of the Earth. Earlier, we developed the technique for determining (in quadratic approximation)
the contribution of anomalous masses distributed over
ellipsoidal surfaces to a planet’s gravitational field \cite{Ch_2006}--\cite{Ch_2007}
and applied this technique when studying the structure of the envelopes of Earth and Mars \cite{Ch_2010,Ch_2011}. In this paper, we estimate the possible distributions of relief compensation depths and anomalies of density and stresses more
accurately, as well as  anomalous gravitational field in the crust, mantle, and core of Mars. In their similar studies, other authors considered the
compensation of relief masses only at one level, namely, at the Mohorovi\`ci\`c surface ${\bf M}$ \cite{Yuan}--\cite{Turc}. Our
results that were obtained for the Earth show that several compensation levels can exist in the interior of a planet, which agrees with the analysis of the Earth’s
free oscillations and seismological data \cite{Ch_2010}. The compensation depths for different relief harmonics depend strongly on the degree and order of the harmonics. So, our first task was to determine the possible compensation depths for the harmonics of different degrees and orders of the expansion of the Martian topographic heights  with respect to the hydrostatic ellipsoid,

\qquad We solved this problem for the crust and mantle of Mars without allowing for possible stresses in the lithosphere \cite{Ch_2011}. However, our further studies, along with the studies of other researchers \cite{Zarkov_91}--\cite{Ryzg}, showed that the lithospheric layer of Mars is able to endure
appreciable nonhydrostatic loads. Stresses arise that are isostatically unbalanced; ignoring these stresses can yield unreasonably great depths of isostatic compensation for small (in length) relief inhomogeneities
characterized by the high degree harmonics. Since
every relief inhomogeneity is characterized by a specific set of harmonics, the maximum concentration the compensates this set within a specified range of
depths can indicate the most probable compensation depths for the relief inhomogeneity of interest. The compensation depths determined this way will allow
us to find the depth distribution of compensation masses and to determine possible anomalies of the internal gravitational field.

\qquad Mars belongs to the group of terrestrial planets that
have some features in common concerning their formation and composition \cite{Zarkov_91}. The technique we developed for the Earth can be reliably applied to Mars because the results obtained for Earth agree well with
the seismological data and the analysis of free oscillations \cite{Ch_2010}. We must bear in mind, however, that this is a
mere simulation that can be helpful in future studies
on the surface of Mars.

\section{ Compensation depths for relief masses  }

\qquad The solution to the problem must satisfy a system
of two equations; one equation reflects the correspondence between the observations and the contribution of the relief masses and compensation masses to the
gravitational field, while the other is the condition of
the pressures below the compensation depth, which is equal to the hydrostatic pressure. The compensation depth $ d_{nm} $ for an arbitrary relief harmonic $ a_{nm} $ is obtained from the relationship
$$
d_{nm}=R_0-R_M \left(  a_{nm}^{M1}/b_{nm}^{M2}   \right)^{1/n},
\eqno (1)
$$
where $R_0=3389.5$~km is the mean radius of Mars;
and  $ a_{nm}^{M1} $, $ a_{nm}^{M2} $ are the coefficients in the spherical function
expansion of the altitudes of compensation surfaces
${\bf M1}$ (obtained using the gravitational field after deduction of the contribution of relief masses in the quadratic approximation) and ${\bf M2}$ (obtained using the
hypothesis of isostatic compensation for the relief
masses) with respect to the corresponding hydrostatic
ellipsoid, with the compensation radius $R_M$ being fixed
(a similar relationship is valid for $ b_{nm} $ when $ b_{nm} $ is substituted for $ a_{nm} $).

\qquad As can be seen, solution (1) is possible, i.e.   $ d_{nm}\geq 0 $, if  $ 0\leq a_{nm}^{M1}/b_{nm}^{M2}\leq (R_{0}/R_M)^n$.
Studying solution (1) for both Earth and Mars shows that it is possible to
compensate the relief masses at one level only for a certain set of harmonics. To make the compensation for the remaining harmonics, when the solution does
not correspond to the obtained condition for $ d_{nm}<R_0 $, the
following two versions were chosen to ensure the minimum deviations between the internal composition of Mars and the equilibrium state:

(1) Compensation is made at two levels. The upper
crust is taken as the first level. The possible depths of
the second level are determined from the analysis of
the results that were obtained for the harmonics for
which solution (1) exists. The final choice is made
with allowance for the weight function, which is
inversely proportional to the deviations of the internal
composition of Mars from the hydrostatic equilibrium.

(2) The uncompensated relief harmonics yield the stresses in the lithosphere of Mars provided that the stresses do not exceed the strength limit  of the lithosphere.

\qquad We used the harmonic expansion of degree $n \leq 18$  as
the initial presentation for the gravitational field of
Mars \cite{Konopliv} and for the relief \cite{Zuber} after taking the hydrostatic values into account \cite{Zarkov_93}. Figure 1a shows the
distribution of the mean compensation depths for the relief harmonics as function of degree $n$ of the harmonic. The mean depth was calculated separately
when considering the compensation in the upper crust
($d \leq 25$ km). We did not analyze the harmonics whose amplitudes were smaller than the mean amplitude for the considered degree and the deviation of the compensation depth from the mean value exceeded  3$\sigma$.
These harmonics are likely to contribute to the stresses
in the lithosphere of Mars. Figure 1 shows that the
relief harmonics of low degrees $n=2$ $(a_{20}, a_{22}, b_{22})$, $n=3$ $(a_{33})$ and $n=4$ $(b_{44})$  are compensated at depths of $d=$1000--1350 km. The large scatter in compensation
depths for the harmonics of degrees $n=3,4$ is due to the harmonics 
$b_{32}, a_{40}$ ($d=$320~km), $b_{33}$ ($d=$450~km), $a_{32}, b_{42}$ ($d=$760~km), $a_{33}$ ($d=$1250~km), and $b_{44}$ ($d=$1050~km).
 The harmonics of degrees $n=$5--18 are
compensated mostly at depths of 34–-240~km; the scatter relative to the mean values is small, with the exception of $a_{55}$ ($d=$400~km) and $a_{53}$ ($d = $270~km).

\qquad As for the harmonics that do not satisfy condition (1),
only two harmonics ($a_{30}, b_{31}$) can be isostatically compensated according to the first version as follows from the analysis of harmonic amplitudes of the simple layer corresponding to the relief and upper crust; the other
harmonics lead to stresses in the lithosphere of Mars.
Figure 1b shows that the relative contribution of stresses achieves the maximum for the harmonics of degrees  $n \geq 11$ for which the load pressure of the relief
and anomalous masses of the crust is compensated mostly due to stresses (according to \cite{Koshl_Zarkov}, the contribution of these harmonics is greatest for the Olympus Mons). For the same harmonics, the correlation coefficient is greatest between the load pressure and stresses ($k =$ 0.72-–0.99), which implies that Olympus creates mainly compressive stresses (Fig. 2b). The correlation coefficient is lowest for the harmonics of low
degrees ($n\leq 7$), namely, $k <.05$ for $n =$ 1-–3 and
$k < 0.3$ for $n =$ 4-–7; this indicates that large relief inhomogeneities spanning more than  25\textdegree(Tharsis, Hellas, Argyre) are mostly isostatically compensated or partly overcompensated (the correlation between
the stresses and relief is negative for the harmonics of degrees $n =$ 1-–4, 6, 7).

\begin{figure}[t]
\includegraphics[width=0.99\textwidth ]{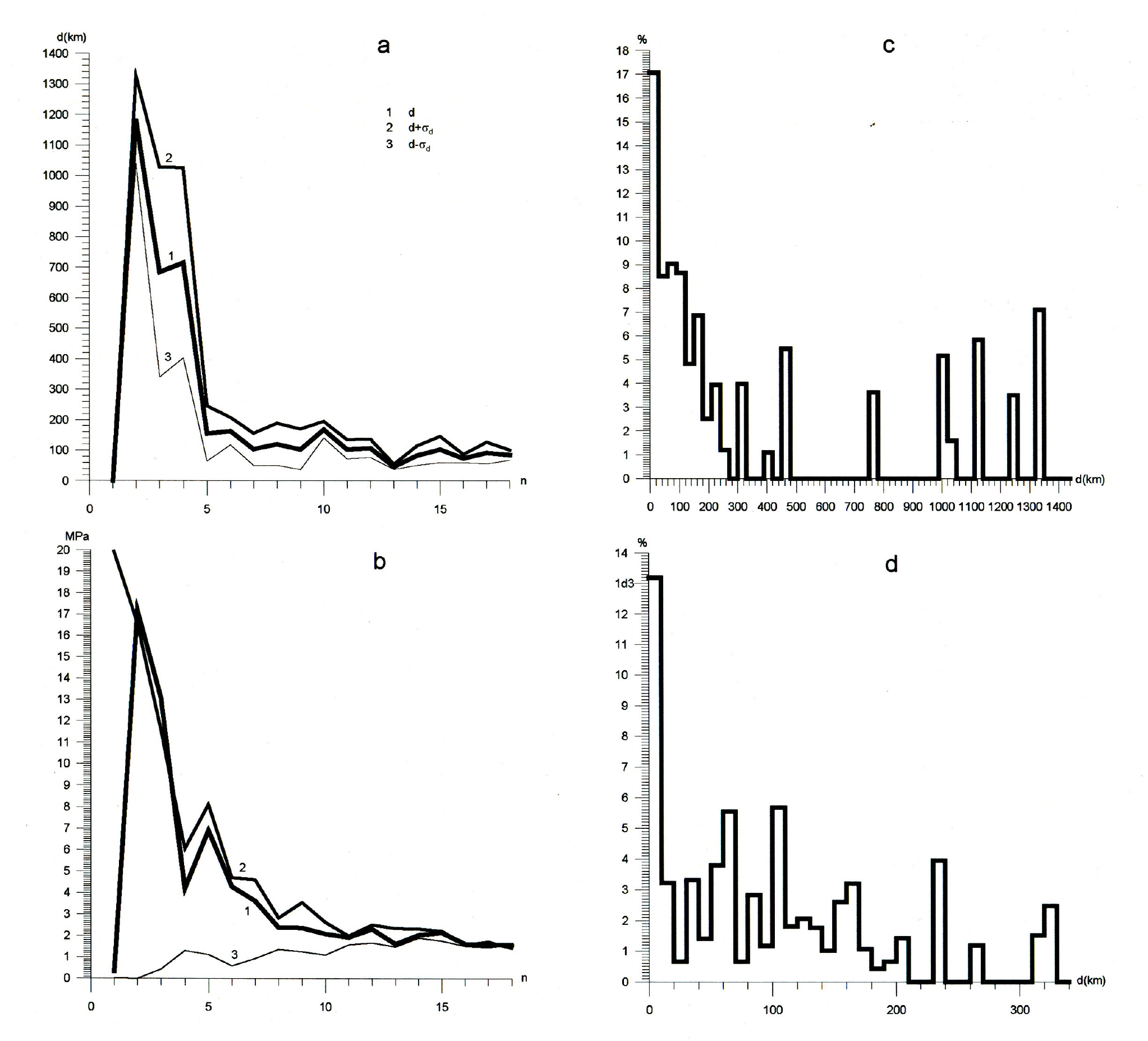}
\hfill
\centering
\caption
{ \footnotesize 
Fig.~1.
(a) Distribution of the compensation depths of relief masses as a function of degree $n$ of the expansion; (b) distributions of
the loading pressure of homogeneous relief (2), relief and anomalous masses of upper crust (1), and stresses (3) as functions of
degree $n$ of the expansion; (c) distribution histogram of the compensation depths for the relief harmonics with a step of 30 km for
the entire depth range between 0 and 1400~km, (d) step of 10 km for the depth range between 0 and 340~km.}
\label{Fig1}
\end{figure}

\qquad The histograms in Figs. 1c, d show the distributions of compensation depths for the relief harmonics after excluding the harmonics responsible for the stresses.
The histograms and mean depths were calculated with account of the weights corresponding to the contribution of the harmonics to the gravitational field. The
analysis of the histograms allows us to conclude that the relief is completely compensated in the depth range $d$ = 0--–1400~km, with more than 17\% of compensation occur in the upper crust ($d=$0---25~km,$\bar d  =5.0\pm 3.6$~km). Further, several major compensation layers can be singled out: a crust–mantle
transition layer ($d =$ 30--–210~km, $\bar d   =100\pm43$~km); a
lithospheric boundary layer ($d =$ 230--–330~km, $\bar d =280\pm 41$~km); an upper–middle mantle transition layer ($d =$ 390--–460~km, $\bar d = 450 \pm 11$~km); an anomalous layer in the middle mantle ($d =$ 750--–770~km, $\bar d = 760 \pm 5$~km); and a mantle–core transition layer ($d =$ 1000--–1350~ km, $\bar d = 1180 \pm 138$~km). When
applied to the Earth, this technique shows that the
most probable compensation depths agree well with
the depths based on the seismic data and spectral analysis of the normal modes of Earth’s free oscillations.
Applying this technique to Mars will most likely yield
reliable results.

\section{Stresses and density anomalies in the crust and mantle of Mars }

\begin{figure}[t]
\includegraphics[width=0.99\textwidth ]{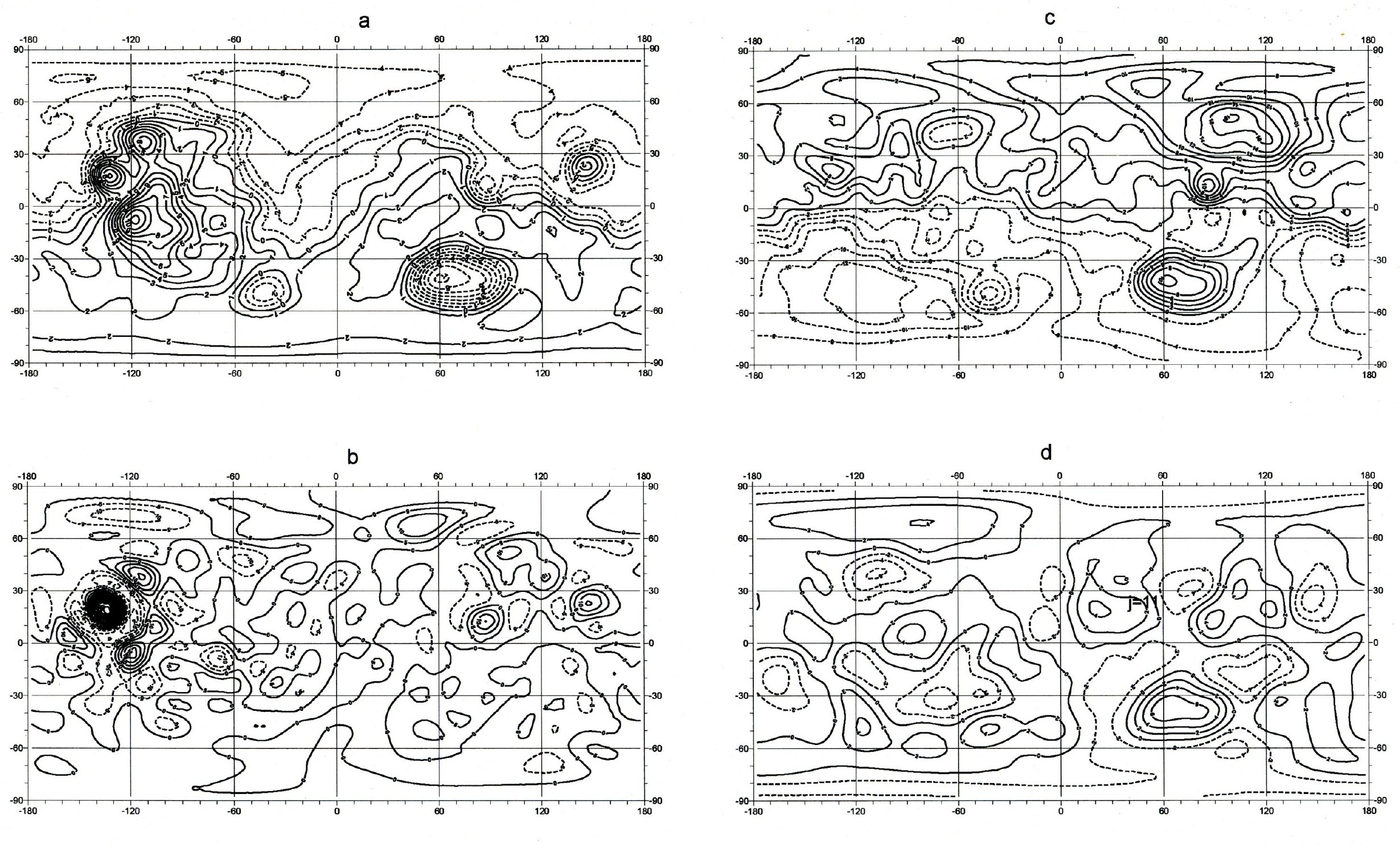}
\hfill
\centering
\caption
{ \footnotesize 
Fig.~2. Anomalous structures of the Martian crust for the expansion up to the 18th degree: (a) --  relief heights relative to the hydrostatic ellipsoid, the cross section of isolines is 1 km; (b) --  anomalies of the vertical stresses in the crust, the cross section of isolines
is 5 MPa; (c) --  anomalous masses of the upper crust (at a depth of 0–25~km), the cross section of isolines is $2 \cdot 10^6$ kg/m$^2$;
(d) --  anomalous masses of the crust–mantle transition layer (at a depth of 30-–210 km), the cross section of isolines is $2 \cdot 10^6$ kg/m$^2$ }
\label{Fig2}
\end{figure}

\qquad Figure 2a shows the smoothed relief of Mars presented in the form of an altitude expansion with respect to the hydrostatic ellipsoid for $n \leq 18$. Here, the major relief features on Mars that are greater than 10\textdegree
are reflected. The distribution of possible stresses in the lithosphere of Mars is mapped in Fig. 2b. It can be seen that the major vertical compressive stresses (positive values) correspond to the volcanic craters (Olympus, Arsia, Pavonis, and Elysium Montes), Alba Patera, Utopia and Isidis Planitie, and the eastern part of
Vastitas Borealis. They are encircled with the vertical tensile stresses (negative values). The tensile stresses also correspond to the Valles Marineris, Echus Chasma, and the western part of Vastitas Borealis. All the stresses do not exceed the compressive strength (266 MPa) and tensile strength (22 MPa) of basicrocks \cite{Ryzg}. Since the strengths of acid intrusive rocks (granites) are twice as low, the large patterns of faults
surrounding Tharsis and flanks at the Vallis Marineris
are likely to indicate the more acidic composition of rocks in the regions of faults. They correspond as well to negative anomalies of crust density (Fig. 2c). The Argyre Planitia and Hellas Planitia, as well as Planum Boreum and Planum Australe, seem to be mostly compensated and do not create large stresses.

\qquad Figures 2c, d and 3a, b, c, d present the maps of
density distributions of compensation masses recalculated relative to the density of the simple layer at middle depths of 0, 100, 280, 450, 760, and 1180~km,
which correspond to the set of harmonics for the layers
with above middle depths. The comparison of these density distributions allows us to draw the following conclusions:

\qquad(1) The dichotomy of the relief on Mars caused by the harmonics of the first degree is mostly compensated due to lava filling of the crust in the plains of the
northern hemisphere (Fig.~2c). The impact features (Hellas, Isidis, Utopia, Argyre), which are possibly similar to the lunar mascons of enhanced density, have
roots stretching down to depths of $280 \pm 41$~km and are
surrounded by circular features of reduced density(Figs.~2c, d; 3a).

\begin{figure}[t]
\includegraphics[width=0.99\textwidth  ]{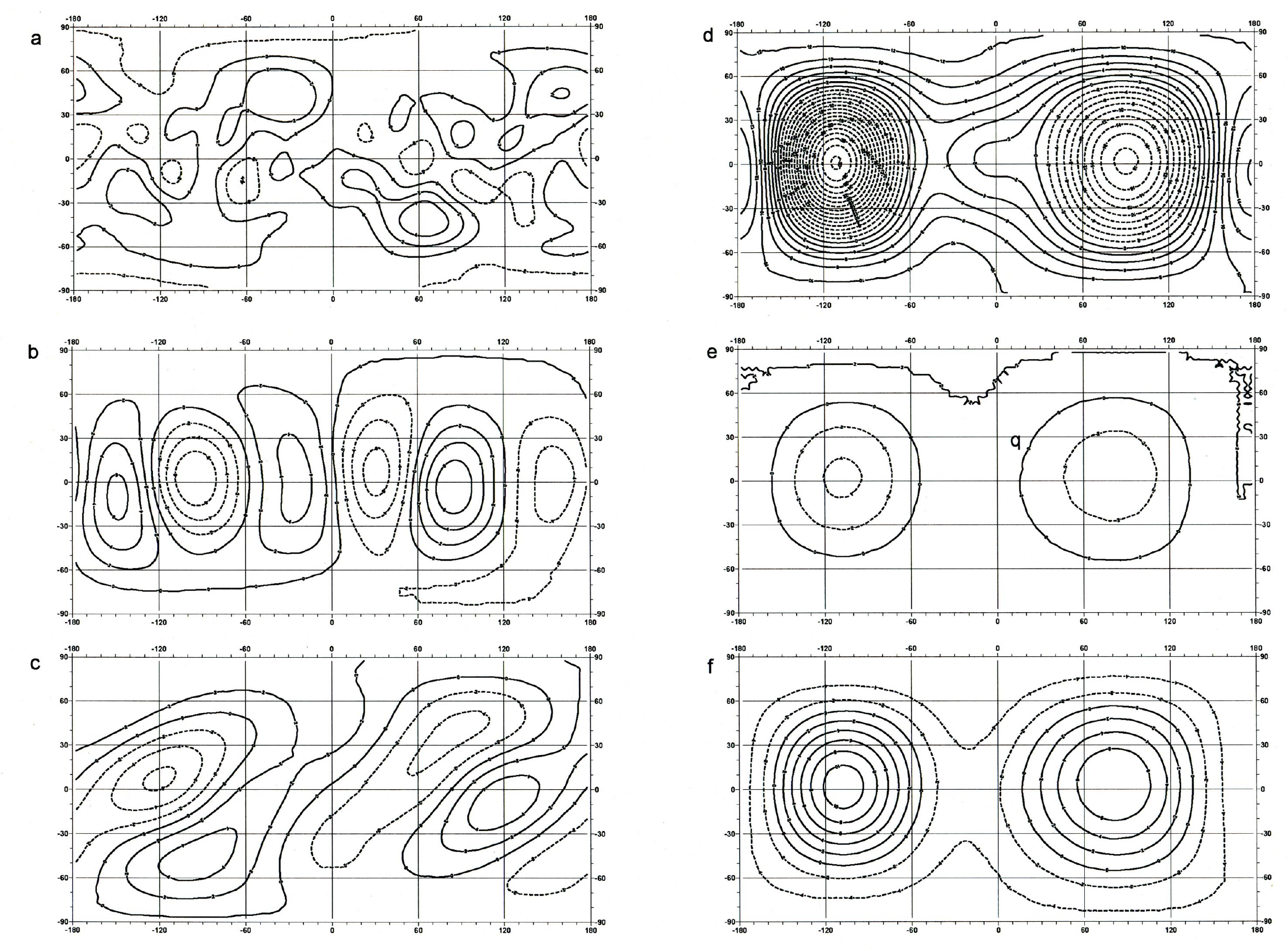}
\hfill
\centering
\caption
{ \footnotesize
Fig.~3. Anomalous masses of the mantle and core of Mars for the expansion up to the 18th degree, the cross section of isolines is $2\cdot 10^6$~kg/m$^2$: (a)--- at a depth of 230--330~km;
(b) --- at a depth of 390--460~km; (c)--- at a depth of 750-–770~km; (d) --- at a depth of 1000–1350~km; (e)--- at the outer core boundary; 
(f)--- at the inner core boundary.
}
\label{Fig3}
\end{figure}

\qquad (2) There is no clear correlation between the possible depths ${\bf M}$ that were obtained using Fig. 2d by taking the density jump at ${\bf M}$ into account as well as the relief features (with limited exceptions for Hellas and
its surroundings, as well as Utopia, Elysium, Argyre,
Arsia, and Alba, due to local compensation at ${\bf M}$). The
boundary at ${\bf M}$ could have been raised due to flows
ascending from the mantle below Vastitas Boreallis,
Utopia, and Arabia Terra (Fig. 3a). The regions of
negative anomalies (including circumpolar ones) are
likely to indicate fluid supplies (perhaps water) that
did not have enough time to reach the Martian surface
through the solidified crust.

\qquad (3) The relief anomalies of the Tharsis volcanic plateau and symmetric feature in the eastern hemisphere,which are mostly caused by the harmonics of the second degree, seem to have appeared and been maintained dynamically due to two plumes in the mantle substance enriched with fluids. They originate in the
lower mantle (Fig.~3d) and divide into several branches in the upper mantle (Figs.~3a, 2d).

\qquad (4) Figure~3b shows that the ascending flow of a
light substance in the mantle, which is stronger below Tharsis and pushes the descending heavier masses aside, while the stronger descending flow in the symmetric equatorial region pushes the light ascending masses aside. The sloping features in Fig.~3c are also of interest. They may arise due to the lateral gradient of
gravitational potential (Fig.~5c) and the Coriolis force.
The heavier components of the mantle substance that descend from a depth of 450~km shift to the South (toward the maximum potential) and then, affected by
the Coriolis force, to the West, as well as North, and
then East. The lighter components ascending from a depth of 1180~km shift toward the minimum potential, namely, to the North and then East.

\section{The sources of plumes and the forces that maintain their existence }

\qquad  Obtained by us the density anomalous in the crust and mantle of Mars  force the internal gravitational field to deviate appreciably from the hydrostatic equilibrium in the core, although they satisfy the condition, thus suggesting that the pressure is hydrostatic below 1350~km; this may cause convective motions to
appear. The fact that Mars does not have an intrinsic magnetic field can indicate either the hydrostatic equilibrium of the core or possible convective motions
that are symmetrical relative to the polar axis (which
does not satisfy the operating condition of a hydromagnetic dynamo). 
To verify these conditions, we considered two three-layer models that do not distort
the external gravitational field of Mars and satisfy the condition of hydrostatic equilibrium below the lower layer. The first model suggests that the core as a whole
is in hydrostatic equilibrium. To meet this model condition, the lower layer must correspond to the boundary of the external core. 
With the density jump at the
core boundary being $3 \cdot 10^3$ kg/m$^3$, the anomalies in
density can correspond to variations (–6, 13)~km for the accepted core boundary ($R = $1700~km); a part of
these variations, namely, (–6, 3)~km can be accounted for by the fact that the core compression exceeds the hydrostatic compression (0.0053) by 0.0027. Such
excess in the core compression over the hydrostatic compression is unlikely to correspond to the modern liquid core.
 Thus, we considered another model,
whose lower layer corresponds to the accepted boundary of the solid inner core ($R=1000\pm 100$~km) while
the two upper layers correspond to the external core
boundary ($R=1700\pm 300$~km) and the transition layer between the lower mantle and core ($d = $1000---1350~km). The distributions of anomalous masses obtained in
this core model are presented in Figs.~ 3e,~f.

\qquad 
We mapped the relative anomalies of internal gravity (relative to $g_0=$3.73 m/s$^2$) using the obtained distributions of anomalous masses (Fig.~4). The maps
outline the regions where the differentiation of the light and heavy components of the Martian material occurs with greater rates in the areas of positive anomalies. Thus, Fig.~4a shows that the ascending plumes of light substance originate from the equatorial regions of the core where the maximum rate corresponds to the core regions below Tharsis and the diametrically opposed region.
 The pattern of anomalies presented in Fig.~4a holds within the region between the boundaries of the inner core and up to the transition mantle–core layer. The pattern of anomalies changes above this layer (Fig.~4b). The differentiation rate achieves a minimum at the plume area, which leads to the trapping of light matter at the plume area spanning up to the lithosphere boundary (Fig.~4d).
 The differentiation rate changes in the lithosphere (Fig.~4e,~f) according to the plume fragmentation into separate branches. Figures~4e,~f show that the minimum differentiation rate corresponds to the highest and isostatically uncompensated relief features (Olympus, Arsia,
Pavonis, and Elysium Montes) while the regions of
positive anomalies correspond to lowlands compensated in the lithosphere and upper mantle (Hellas Planitia, Vastitas Borealis, and Acidalia Planitia).

\section{Convective motions in the mantle and core of Mars}

\begin{figure}[b]
\includegraphics[width=0.99\textwidth ]{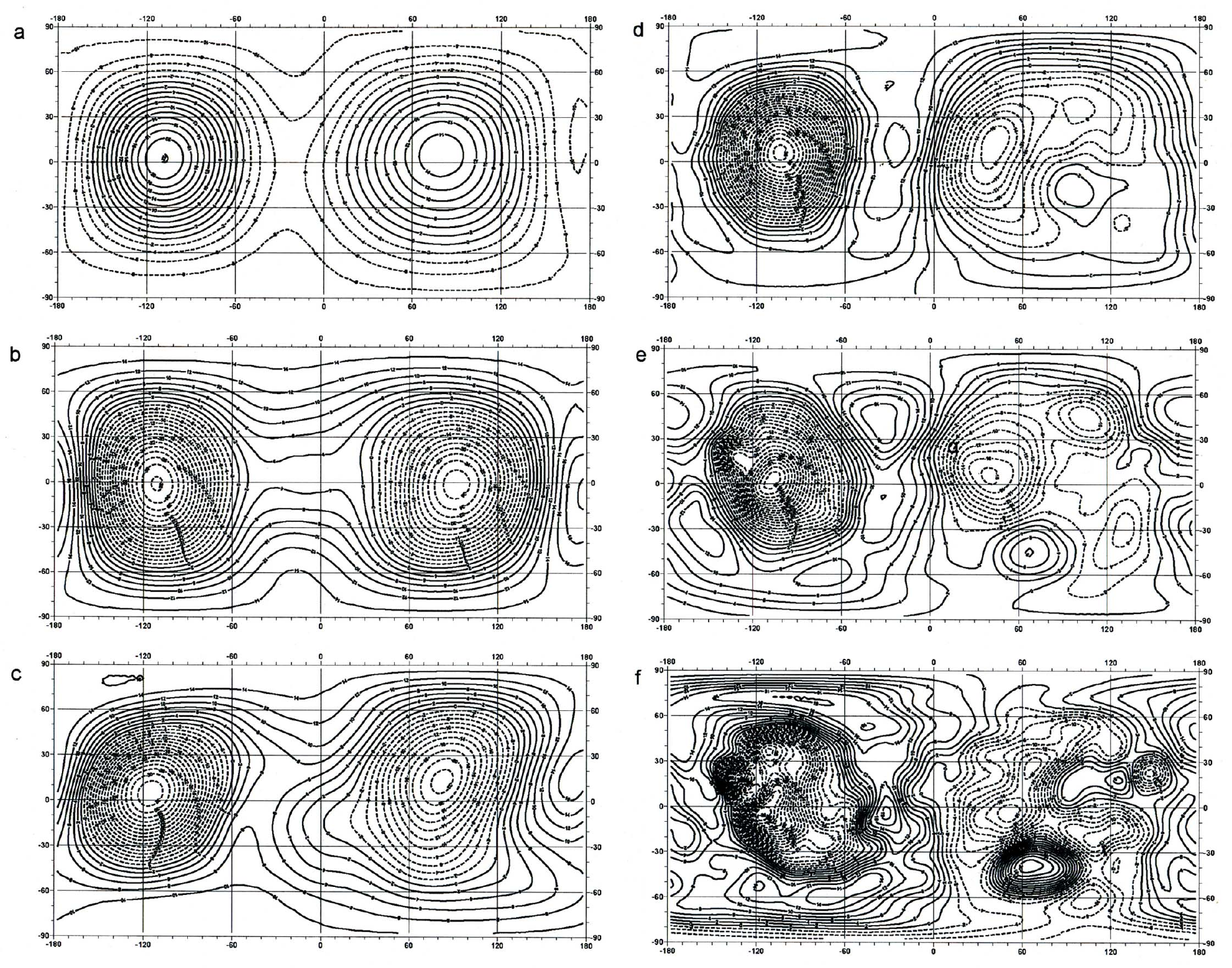}
\hfill
\centering
\caption
{ \footnotesize 
Fig.~4. Relative anomalies of the attractive force in the core, mantle, and crust of Mars; the cross section of isolines is $2\cdot 10^{-4}$ $g_0$:
(a)--- above the inner core boundary (for $R = $1100~km); (b)--- above the mantle–core transition layer (for $R = $2400~km); (c)--- above the
layer at a depth of 760 km (for $R = 2700$~km); (d)--- above the upper–middle mantle transition layer (for $R = 3000$~km); (e)--- abov
the lithosphere boundary layer (for $R = 3170$~km); (f)--- above the crust–mantle transition layer (for $R = $3360~km)
}
\label{Fig4}
\end{figure}

\qquad Convective motions can appear if forces exist that
cause lateral motions of matter. In the gaseous and liquid envelopes of the planets (and Sun), the Coriolis
forces that arise due to radial motions in the spherical
envelopes can act in the lateral direction. In the solid
envelopes, the lateral gradients of gravitational potential can be such forces; as a result, the heavy components not only move down (which is energetically
expedient) but also shift laterally toward the maximum
potential, while the light plume components move
toward the minimum potential.

\begin{figure}[t]
\includegraphics[width=0.99\textwidth ]{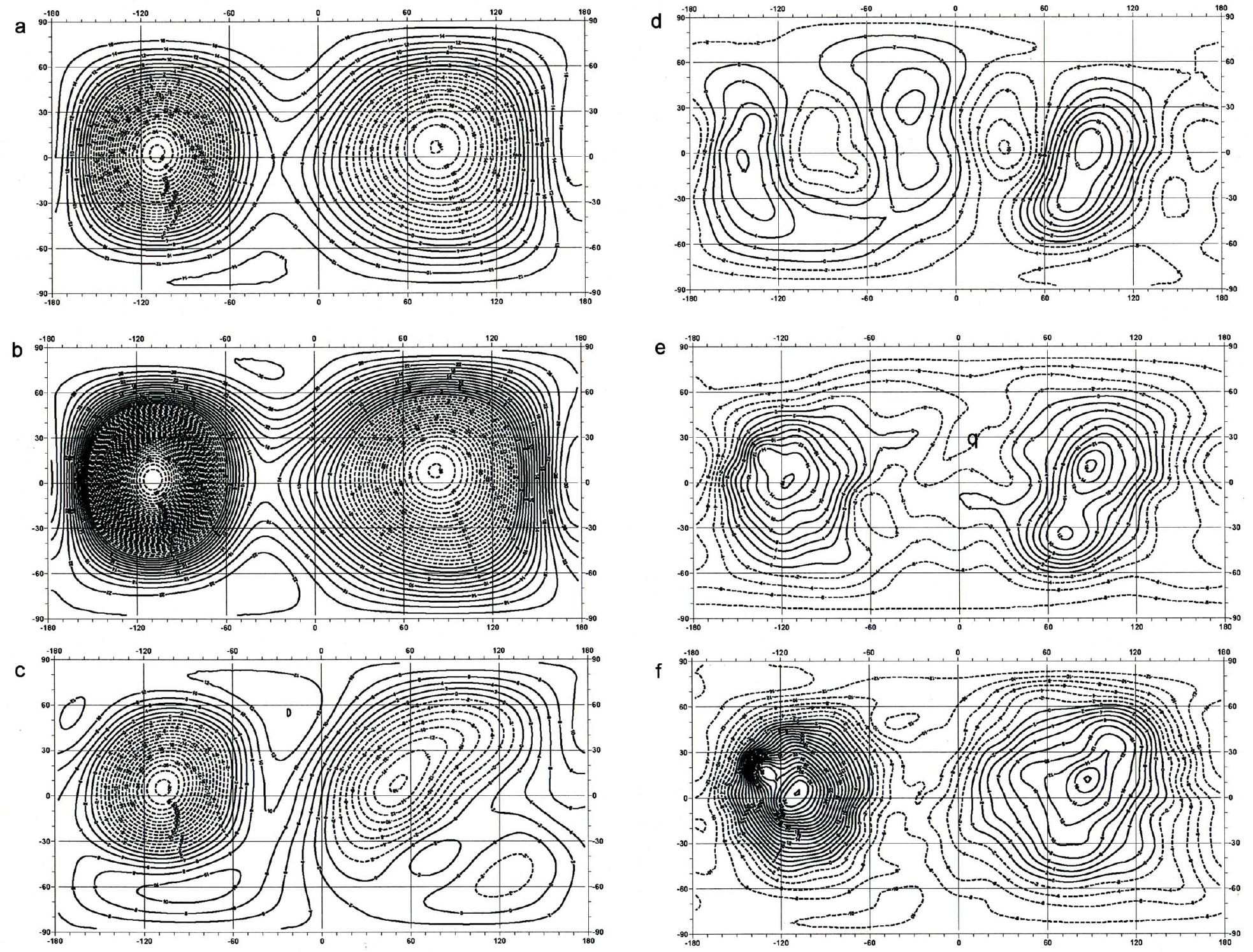}
\hfill
\centering
\caption
{ \footnotesize
Fig.~5. Relative anomalies of the attractive potential in the core, mantle, and crust of Mars; the cross section of isolines is $2\cdot 10^{-5} V_0$:
(a)--- at the outer core boundary; (b)--- at a depth of 1180~km; (c)--- at a depth of 760~km; (d)--- at a depth of 450~km; (e)--- at a depth of 280~km; (f)--- at a depth of 100~km.
 }
\label{Fig5}
\end{figure}

\qquad The lateral distributions of the anomalies of inner
potential caused by anomalous masses (Figs.~2,~3) at
various depths with respect to $V_0 = g_0 R_0$ are mapped in Fig.~5. The maps are presented upward from the outer core boundaries because the distribution pattern inside the core is similar and the anomalies rapidly decrease to zero at the inner core boundary. 
Figures~5a,~b,~c show
that the lateral gradients of potential achieve the maximum in the mantle–core transition layer and that the behavior of their distribution scarcely changes until
the upper–middle mantle transition layer. It is interesting to note that the distribution of potential at depths of 450, 760, and 1180~km correlates with the
distribution of anomalous masses at these depths (Figs.~3b,~c,~d). This signifies that the distribution of anomalous masses at these depths is stable against the
gravitational effect of anomalous masses at other levels
and is defined by the potential of these masses. The situation is entirely different at depths of 280, 100, and 30~km, where the distribution of potential is due to the
masses of the relief and upper crust. Thus, the anomalous features at these depths (Figs.~2d,~3a) are unstable and depend on processes that change the structure of
relief and upper crust.

\qquad As for the convective motions due to variations of
potential, these motions can cover the entire mantle between the core and lithosphere boundaries (where the lateral motions are directed opposite to the
motions near the core boundaries) (Figs.~2–5).
The flows of electrically conductive material that arise in the meridional planes can create toroidal magnetic fields; since the magnetic fields in the northern and
southern hemispheres have opposite directions and are nearly symmetrical, the total field approaches zero. 
Similarly, the weaker flows in the planes parallel
to the equatorial plane are almost symmetric relative to the rotational axis and cannot create a global poloidal magnetic field, according to the Cowling theorem.


\bigskip
{\Large \bf \quad Conclusions}
\medskip

\qquad The crust and mantle of Mars are characterized by
nonuniform distributions of density and stresses down
to a depth of 1450~km.  The relief anomalies of the
Tharsis volcanic plateau and symmetric feature in the
eastern hemisphere are likely to have arisen and been
maintained dynamically due to two plumes of the
mantle substance, which originate at the core boundary and are enriched with fluids. The plumes can arise and be maintained due to anomalies in the inner
attractive forces. The volcanic craters of enhanced density and large relief depressions are maintained
mostly due to the elastic properties of the lithosphere
and create stresses that do not exceed the strength of
the lithosphere.


\end{document}